\documentstyle[multicol,aps,prl,epsf]{revtex}

\begin{document}
\title{Vector lattice model for stresses in granular materials}
\author{Onuttom Narayan}
\address{Physics Department, University of California, Santa Cruz, CA
95064}
\date{\today}
\maketitle
\begin{abstract}
A vector lattice model for stresses in granular materials is proposed.
A two dimensional pile built by pouring from a point is constructed 
numerically according to this model. Remarkably, the pile violates the
Mohr Coulomb stability criterion for granular matter, probably because
of the inherent anisotropy of such poured piles. The numerical results
are also compared to the earlier continuum FPA model and the (scalar)
lattice $q$-model.
\end{abstract}
\begin{multicols}{2}
\narrowtext
Understanding how granular materials sustain externally applied loads
is an interesting and difficult scientific problem, with obvious
engineering implications. Unlike for elastic materials, the slippages 
that occur when a granular heap is built make it impossible to define a 
displacement (and thence a strain) field. One thus has to solve for a 
tensor stress 
field (which cannot be expressed in terms of a vector displacement
field) using the vector force balance equation. In two dimensions, 
where the stress tensor has three independent components, 
this results in one `missing' equation. In three dimensions, the 
situation is even worse.

One of the main tools used to deal with this difficulty is the Mohr
Coulomb (MC) stability criterion. This states that the ratio of the 
shear to the normal stress at any point inside a cohesionless granular 
heap, with any 
orientation of axes, cannot exceed the coefficient of friction $\mu.$
This is natural, since otherwise one would expect an appropriately
located (and oriented) slip plane to spontaneously destabilize the 
heap. Using the MC inequality and symmetry (or further
assumptions), it is sometimes possible to obtain useful results on the 
stability of granular heaps~\cite{Nedder}.
For instance, the elastoplastic model~\cite{Goddard} of granular stresses 
relies crucially on this criterion. The fixed principal axes (FPA) 
model~\cite{Cates1,Cates2} does not invoke the MC criterion, but obtains
it as a byproduct.

{\it Both\/} the elastoplastic and the FPA theory yield the surprising 
result that the MC
criterion is saturated in a large outer region of a two-dimensional 
pile built by pouring from a point. This implies~\cite{NN} that 
an infinitesimal perturbation inside the outer region 
should destabilize the pile, while one would expect
that a poured pile should only be unstable at the surface.

A byproduct of the MC criterion is that the angle of repose
$\phi$ should satisfy the equation 
\begin{equation}
\tan\phi = \mu.
\label{anglerepose}
\end{equation}
In two dimensions, the MC criterion becomes
\begin{equation}
\label{M-C}
{{(\sigma_{xx}-\sigma_{yy})^2+4\sigma_{xy}^2}\over
{(\sigma_{xx}+\sigma_{yy})^2}}\leq\sin^2\phi.
\end{equation}

Although the MC criterion seems reasonable, there is 
the implicit assumption that the granular medium is 
isotropic. This is a questionable assumption for a granular heap built
by pouring from a point. As the grains 
roll down the slopes of the heap, 
anisotropy can be `frozen in', with different behavior 
parallel and perpendicular to the slope. In fact, for a two dimensional
heap of identical disks in a perfect triangular lattice~\cite{Hong}, 
the left hand side of Eq.(\ref{M-C}) goes to unity at the corner of
an infinitely high heap, while $\phi= \pi/3.$ The motivation for 
this paper is to investigate whether this anisotropy persists
away from the special case of the perfect lattice.

\begin{figure}
\centerline{\epsfxsize \columnwidth \epsfbox{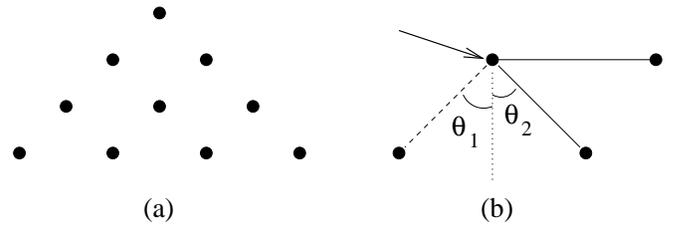}}
\vskip -0.8in
\caption{(a) Triangular lattice. Each site rests on its two neighbors
below, and receives forces from its neighbors above. 
(b) The incoming force is too horizontal, and the site is unstable.
The downwards bond going left is broken, and a horizontal bond is 
established. The angles $\theta_1$ and $\theta_2$ are independent random
variables at each site.}
\label{fig1}
\end{figure}
In this paper, we consider a two dimensional lattice model, with the 
grains placed on a triangular lattice (see Figure~\ref{fig1}). Each grain 
rests on its two 
neighbors in the row below, and supports its neighbors in the 
row above. Adjoining sites in the same horizontal row are not connected.
Forces propagate downwards from each grain to its descendants. In addition
to the force from its predecessors, a grain also transmits its own weight
downwards. This weight is taken to be the same for every grain.

In order to include disorder in the model, the outgoing bond angles 
$\theta_1$ and $\theta_2$ from a site  are chosen randomly. For 
$N$ lattice sites, there are $2N$ random angles, each of which is 
chosen independently from a uniform distribution over the interval
$[\theta_{\min}, \theta_{\max}].$ The width of this interval is a 
measure of the degree of disorder.
With equal sized grains, even if the lattice is viewed as a topological
rather than a geometrical structure, the bond angles would not be all
independent, but such correlations are ignored for simplicity.

Vector force balance is enforced on every lattice site.
Solving for the inter-grain
force network, proceeding downwards from row to 
row, inevitably leads to occasional negative (i.e. attractive) inter-grain
forces. This is unphysical for a cohesionless granular medium.
Accordingly, if the force 
between a grain and one of its descendants is negative, we interpret this 
to mean that the grain is unstable, and rolls in the direction of the 
other descendant. Within the model, this is accommodated by breaking the 
bond with negative force, and establishing a new bond with the 
adjacent site (in the same row as the unstable grain) in the opposite 
direction (see 
Figure~\ref{fig1}). The new bond established necessitates 
recomputing force balance for the sites in the row; several iterations 
can be required before a stable configuration is achieved.

In reality, a moved grain would also change 
the angles for its unbroken bonds. Moreover, since the pile is not really
a regular lattice, the link established to the adjacent site would not be 
horizontal. Both of these issues are ignored. The idea of 
unstable grains moving and readjusting their contacts has been suggested
earlier~\cite{Tkac}, although the implementation
of this process and the specific system studied are different here. 

The model described above is similar to the earlier 
$q$-model~\cite{Science,Copper} with one key difference: 
the $q$-model has {\it scalar\/} forces, 
only keeping track of the vertical force on each grain. Even with this
limitation, the $q$-model successfully accounts for the experimental
result~\cite{Science,Behring,Mueth} that the vertical forces on individual 
grains at the bottom of a granular heap have a distribution $P(f)$ that 
decays exponentially for large $f$. It also agrees with the experimental 
observation~\cite{Mueth} that the 
vertical forces $f_i$ and $f_j$ on two grains at the bottom of a heap
satisfy $\langle f_i f_j\rangle = \langle f_i\rangle\langle f_j \rangle.$
However, the absence of vector forces in the $q$-model prevents it from
reproducing the visually most striking feature of granular forces: the 
existence of ``force chains"~\cite{Dantu,Science}, a sparse network of 
grains that
experience large stresses. Further, while forces on a single grain level
are important for the failure of granular materials, one is often 
interested in a more coarse grained continuum description. There have
been earlier work on vectorizing the $q$-model indirectly~\cite{Cates3}, and
directly with a method different from the one here~\cite{Socolar}. We shall 
see that the vector lattice model proposed in this paper agrees well with 
the same 
experimental results as the $q$-model, and in addition yields force chains,
corroborates some aspects of the FPA model while disagreeing with others,
and permits a test of the MC stability criterion.

The specific system we consider using the vector lattice model is 
that of a granular pile built by pouring from a point. This is because
this is the system on which most of the FPA model analysis has concentrated,
and also because we have reasons to question the validity of the MC
criterion in such a system. For convenience, in the numerical
simulations, the pile is built from the top down instead of from the bottom
up. Thus one starts with one lattice site in the top row and computes
the outgoing forces, proceeding to two sites in the next row and so on. 
If a row has $m$ sites, the next row has $m + 1$ sites {\it unless\/} 
the sites at the edge require adjacent sites to establish horizontal
bonds with. All the numerics
in this paper are with bond angles ranging over $[\pi/6, \pi/3],$ unless
specified otherwise.

\begin{figure}
\centerline{\epsfxsize \columnwidth \epsfbox{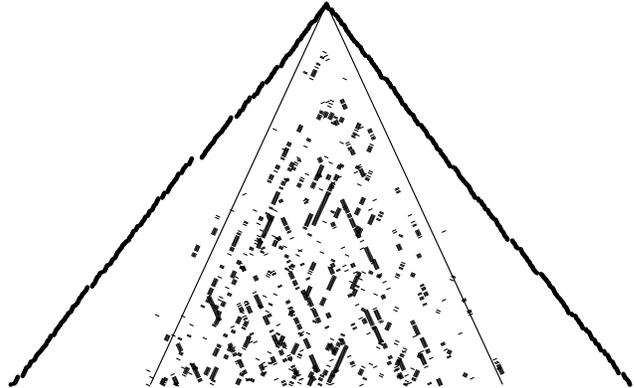}}
\vskip 0.5in
\caption{A pile 400 rows deep. The outer lines are the boundaries of
the pile, and show that it has a well defined slope. In any row, bonds
with a force greater than 4 times the average bond force {\it in that
row\/} are marked, and show directed force chains. The thin straight lines
are the boundaries of the inner region of the pile. The force chains
are concentrated in the inner region. (If the threshold were 
less than 4, there would be more chains in the outer region.)}
\label{fig2}
\end{figure}
Figure~\ref{fig2} shows the result of such a simulation, with 400 rows.
One first notices that 
there is a well defined slope to the pile. From the manner in 
which the pile is built up, we view the central $m$ sites in the $m$'th
row as the interior of the pile, and the sites that flank this inner 
region as {\it buttresses,\/} necessary to stabilize the pile. This division
into an inner and outer region is in accordance with the FPA model. With
the bonds that support unusually high stresses marked, one also sees
force chains concentrated in the inner region of the pile, in accordance with
experiment~\cite{Howell}. Although
the force chains seem perfectly straight, this is because the sites are on a 
regular lattice. With a more realistic model, where the random bond angles
would be accompanied by an irregular lattice, the chains should meander, 
with a mean orientation of $\langle \theta\rangle.$

If the horizontal and vertical forces across the bottom of the pile 
(averaged over many runs) are plotted, the vertical force $f_y$ has a 
flat peak in the inner region and falls off steeply in the outer region. 
The variation in $f_y$ across the inner region is less than $\pm 5\%.$ The
horizontal force $f_x$ varies linearly across the inner region, peaks
roughly at the boundary between the inner and outer regions, and 
falls off in the outer region. The linear profile for $f_x$ shows that
the pile has a strong tendency to {\it splay\/}.
\begin{figure}
\centerline{\epsfxsize \columnwidth \epsfbox{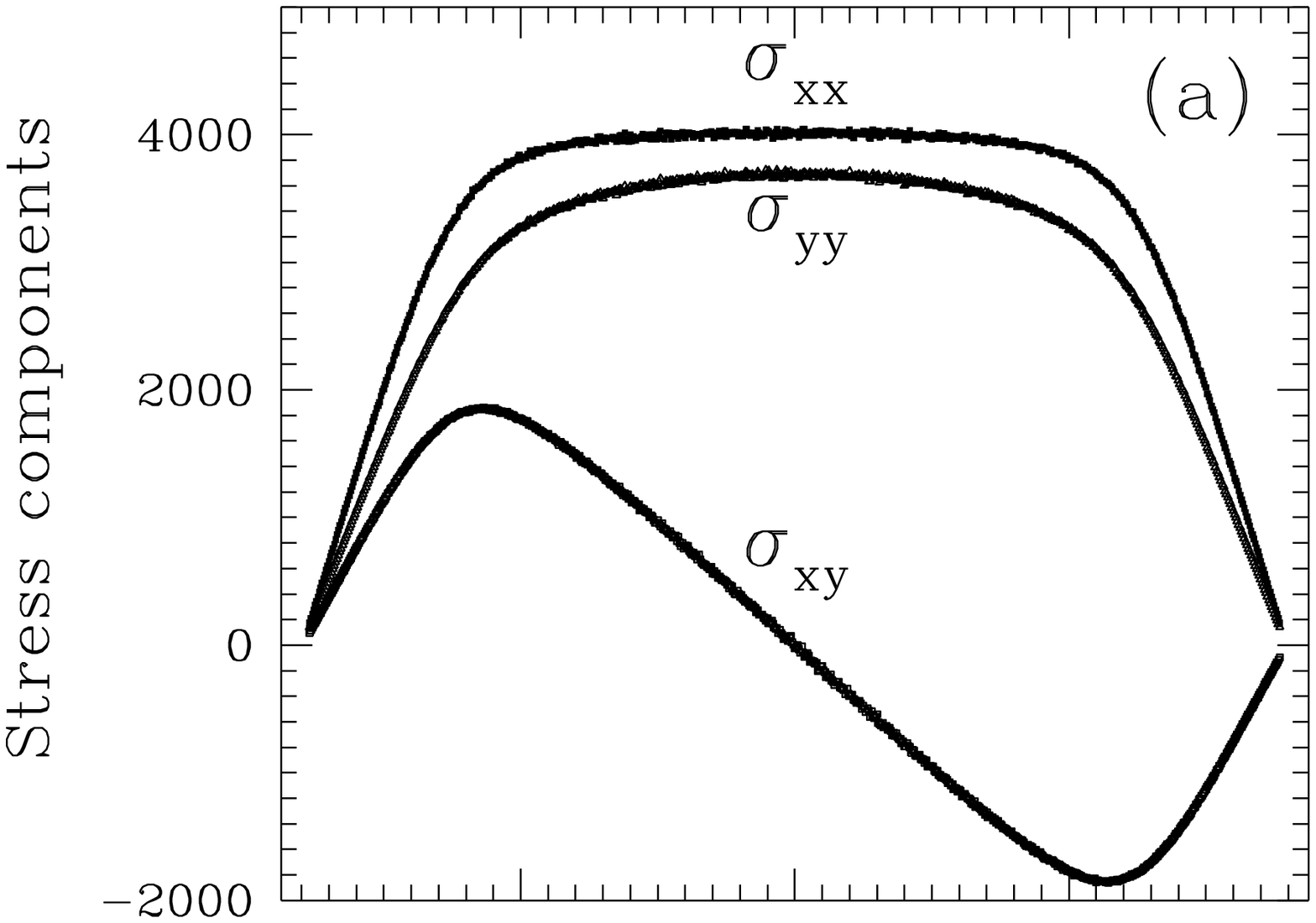}}
\vskip -2.4in
\centerline{\epsfxsize \columnwidth \epsfbox{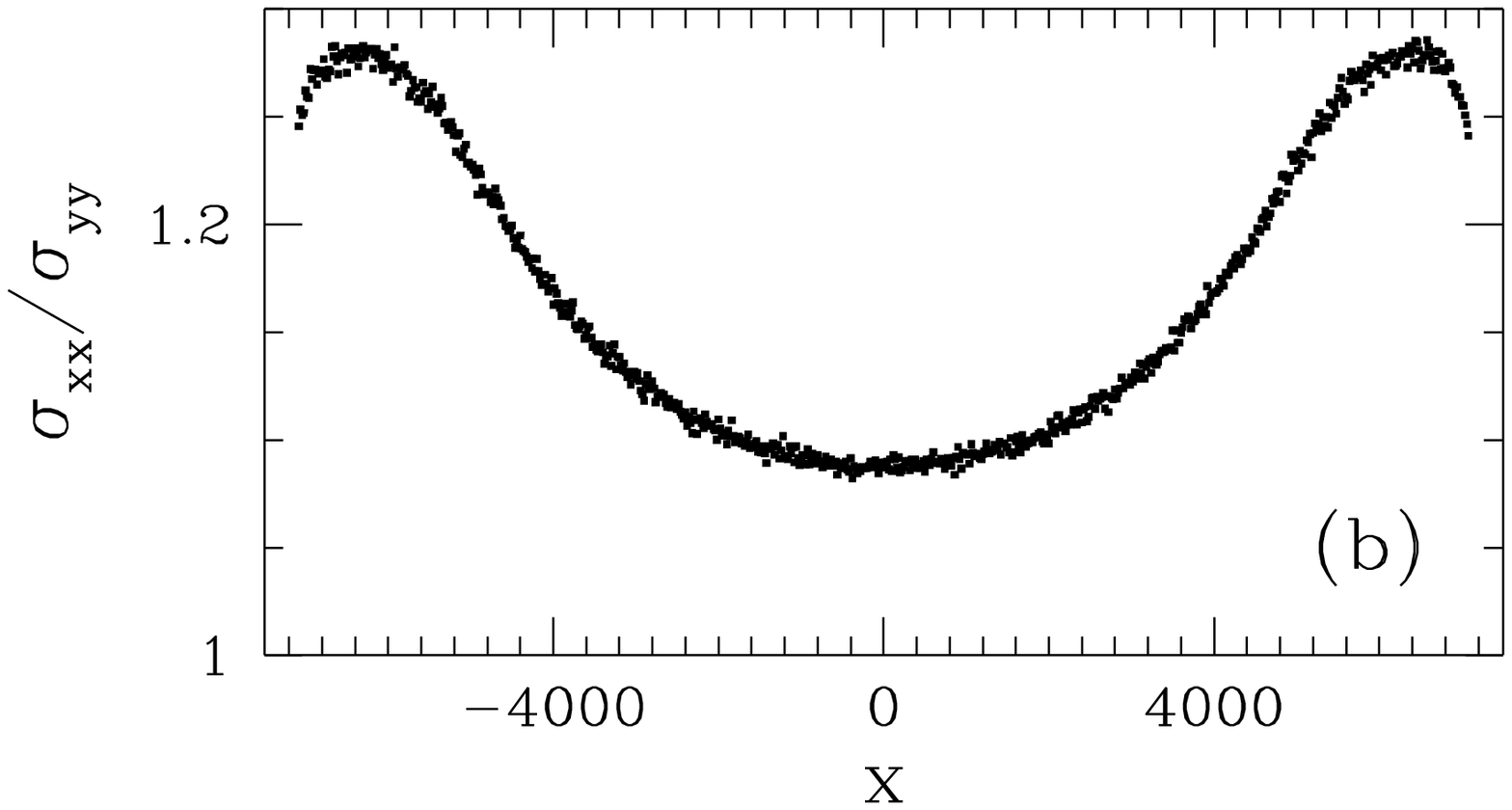}}
\caption{(a) Components of the stress tensor across the bottom of
the pile. The boundary of the inner region is 
at $x=\pm 4000.$
(b) The ratio $\sigma_{xx}/\sigma_{yy}$ from the same data, showing that
$\sigma_{yy}$ is not proportional to $\sigma_{xx}.$
(There is a slight sharpening of the peaks with increasing
pile size.) $\sigma_{xx}$ cannot be expressed even as a linear combination of
$\sigma_{yy}$ and $|\sigma_{xy}|.$}
\label{fig5}
\end{figure}
One can coarse
grain and compute `continuum' stresses: Figure~\ref{fig5} shows the components
of the stress tensor, calculated by coarse graining over twenty lattice sites
at the bottom of a pile of 8000 rows, and then averaging over 2000 runs.
Not surprisingly, the division between
the inner and outer regions seen in the force components
$f_x$ and $f_y$ is also seen here.

The slight curvature seen in $\sigma_{yy}$ in the central region does not
decrease as a function of the size of the pile. This is accentuated in 
the plot of $\sigma_{xx}/\sigma_{yy}.$ The steady increase in this 
stress ratio as one moves away from the symmetry axis 
violates the BCC hypothesis~\cite{Cates1} of $\sigma_{yy}\propto\sigma_{xx}$; 
this cannot be cured by using the FPA or OSL hypotheses~\cite{Cates2} 
of $\sigma_{xx} = k_1 \sigma_{yy} + k _2 |\sigma_{xy}|, $ since 
$\sigma_{xx}/\sigma_{yy}$ is quadratic at the 
center of the pile, whereas $|\sigma_{xy}|$ has a linear kink. More 
importantly, the violation of the FPA hypothesis can be understood
{\it physically.\/} At least in the inner region, the splay in the pile makes 
it increasingly likely for sites to be destabilized outwards as one 
moves away from the axis of symmetry. Such destabilized sites
establish  bonds with their adjacent sites.
Thus while the forces emerging 
from a site in the central region propagate downwards to its descendants,
there is a steady drift towards more horizontal force propagation as one 
moves outwards. This reorients the principal axes of the stress tensor
and increases $\sigma_{xx}/\sigma_{yy}.$ Changing the range of bond
angles, $[\theta_{\min}, \theta_{\max}]$ has no qualitative effect on the 
results discussed so far: the size of the buttresses flanking the inner
region naturally increases when the bond angle range is increased.

\begin{figure}
\centerline{\epsfxsize \columnwidth \epsfbox{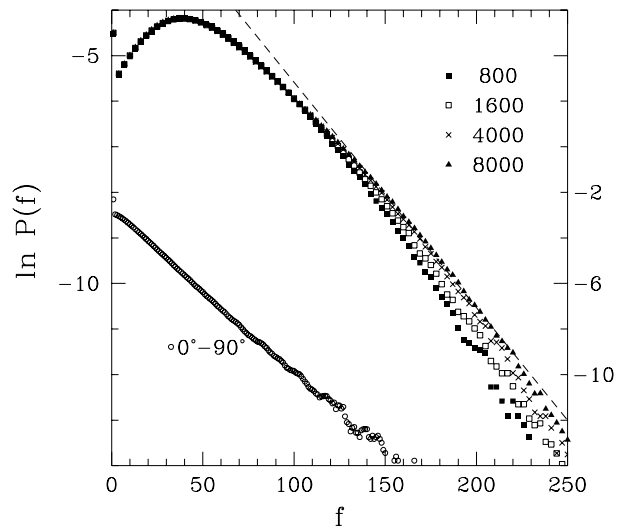}}
\caption{Semi-log plot of $P(f)$ vs $f$ from the inner region of 
piles of varying depths; $f$
is in arbitrary units and scaled with system size.
The asymptote is roughly straight, and grows
straighter with increasing system size. 
The inset (axis label on right) shows a similar plot with bond
angles almost spanning $[0, \pi/2].$}
\label{fig4}
\end{figure}
The approximately
flat profile of the vertical force in the inner region allows us to 
compute the distribution of vertical forces in this region. 
Figure~\ref{fig5} was obtained by
averaging over many runs and lattice sites, but the forces on a single site
in a single run have large fluctuations.
Figure~\ref{fig4} shows a semi-log plot of the probability $P(f)$ for a 
grain in (the inner region of) the bottom row to experience a force $f.$
Although there is a slight deviation from an exponential for large $f,$
this steadily decreases as the size of the pile is 
increased, suggesting an exponential tail to $P(f)$ for infinite system
size. If the range of bond angles is reduced, the evolution towards an 
exponential tail for $P(f)$ with increasing system size is 
slower, but is nevertheless still manifest. Conversely, the inset to 
Figure~\ref{fig4} shows $\ln P(f)$ with a bond angle range of 
$[0.01 \pi/3, 1.49 \pi/3].$ Even though the pile is only 400 rows deep,
$P(f)$ clearly has an exponential tail. 

\begin{figure}
\vskip -0.5in
\centerline{\epsfxsize \columnwidth \epsfbox{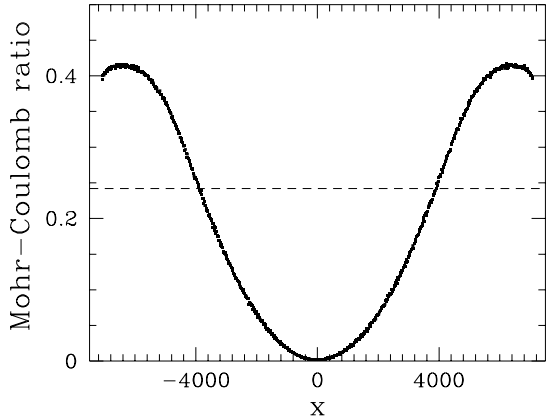}}
\caption{Mohr Coulomb ratio across
the bottom of the pile. The components of the stress tensor are calculated
as in Fig.~\ref{fig5}. The dashed line is $\sin^2\alpha,$ where $\alpha$
is the angle of repose. The curve clearly goes above the line, showing that
Eq.(\ref{M-C}) is violated. There is a slight rounding of the
peaks with increasing pile depth.}
\label{fig6}
\end{figure}
The small $f$ behavior
of $P(f)$ depends on the range of bond angles. However, in all cases
one sees that $P(f\rightarrow 0)\neq 0.$ This is in accordance with 
experiments~\cite{Mueth}, but {\it not\/} with the $q$-model. 
In the model in this paper, if a 
lattice site receives a large force from above, it transmits large
forces to both its descendants. Since it is unlikely that either of 
its descendants gets a large force from its other ancestor, {\it both \/}
descendants are usually destabilized outwards. 
The common descendant of the two destabilized grains thus receives no
force from either of its ancestors. This yields a $\delta$-function in $P(f)$
at $f=0$ (more precisely, at $f$ equal to the weight of a single grain),
which leads to $P(f\rightarrow 0) \neq 0.$ Thus the vector lattice model
agrees with experiments in this respect because it has
{\it arching\/}, which is important in granular 
materials but is absent in the $q$-model. 

One can also compute the correlation between vertical forces on 
adjacent sites (in the inner region of the pile) in the same horizontal row
Unlike the $q$-model this is not exactly zero, and depends on the bond angle 
distribution, but is typically about a few percent of the variance in the 
vertical forces. This is reasonably consistent with the experimental 
observation~\cite{Mueth} that the vertical forces on different grains are
uncorrelated.

Finally, Figure~\ref{fig6} shows the MC ratio, the left hand
side of Eq.(\ref{M-C}), obtained from the coarse-grained stresses of 
Figure~\ref{fig5}. The dashed line shown is at a height of $\sin^2 \alpha,$
where $\alpha$ is the angle of repose. It is slightly tricky to define 
the angle of repose when a regular lattice is combined with random bond 
angles. If the width of the pile at a depth of $m$ rows is $w(m), $ and 
the large $m$ limit of $m/w(m)$ is $\lambda,$ since the average bond angle
is $\pi/4,$ $\alpha$ is taken to be $\tan^{-1} \lambda.$ 
The numerical curve is seen to clearly
go above the straight line, denoting a violation of the MC
criterion. 

In this paper, we have analyzed a new two dimensional lattice model with vector 
forces for granular materials. On an individual grain level, the model yields 
force chains, arching, and is consistent with an exponential tail to 
the distribution of vertical forces. Continuum
stresses show an inner and outer region for a  pile poured from a point,
but the principal axes of the stress tensor
are not fixed. Most importantly, the model 
violates the Mohr Coulomb criterion, calling in question the 
validity of this criterion for poured granular heaps. 

I thank Sid Nagel, Deepak Dhar and Narayanan Menon for useful discussions.

\end{multicols}
\end{document}